\documentclass{ws-procs975x65}
\usepackage{epsfig,amssymb,amsmath,bm,amsfonts,bezier}
\usepackage{graphicx}
\def\c60{C$_{60}$}

\newcommand{\bA}{{\bf A}}

\newcommand{\br}{{\bf{r}}}

\def\be{\begin{equation}}
\def\ee{\end{equation}}
\def\bea{\begin{eqnarray}}
\def\eea{\end{eqnarray}}
\def\bA{{\bf A}}

\def\cH{{\cal H}}

\def\bA{{\bf A}}

\def\br{{\bf r}}

\def\half{{1\over 2}}

\def \cH{{\cal H}}

\def \D02ef{ \Delta_{0\phantom{2}}^{\phantom{0}2}/\epsilon_F }
\def\beq{\begin{equation}}
\def\eeq{\end{equation}}
\def\bea{\begin{eqnarray}}
\def\eea{\end{eqnarray}}
\def\half{\mbox{$1\over2$}}

\def\bA{{\bf A}}

\def\br{{\bf r}}

\def \x2vec{ \boldsymbol{x'} }

\def \omg2vec{ \boldsymbol{\Omega'} }

\def \D02ef{ \Delta_{0\phantom{2}}^{\phantom{0}2}/\epsilon_F }

\def\cH{{\cal H}}

\def\cH{{\cal H}}

\def\sd0{\rho^0_s(T)}

\def\1{\mbox{1\hskip-.25em l}}
\def\6{\langle }
\def\9{\rangle }

\def\sxy{\sigma_{\rm H}}

\begin{document}

\title{Transport Studies of Lattice Bosons:\\
Paradigms for Fluctuating Superconductivity}

\author{Assa Auerbach$^*$ and Netanel Lindner$^\dagger$}

\address{Physics Department, Technion\\
Haifa,32000, Israel\\
$^*$E-mail: assa@physics.technion.ac.il\\
http://physics.technion.ac.il/$\sim$assa/ \\
$^\dagger$E-mail: lindner@tx.technion.ac.il\\
 }

\begin{abstract}
A strong periodic potential  generally enhances  the  short wavelength fluctuations of a superfluid beyond the validity of  standard  continuum approaches.
Here we report some recent results on  hard core bosons on finite lattices. We find several interesting  effects of the  periodic potential on the ground state, vortex dynamics, and  and Hall conductivity.
For example,  the Magnus field on a vortex abruptly reverses direction at half filling.
A rotating Bose condensate on an optical lattice  may allow an experimental  test  of our results.
Insight may also be gained towards about strongly fluctuating  superconductors modelled by
charge  2e  lattice bosons.

\end{abstract}


\bodymatter
\section{Strongly Fluctuating Superconductors}
%\section{Fluctuating Superconductors and Superfluids}\label{sec1}
 ''Conventional'' superconductors undergo a pairing transition at $T_c$, which can be well described by BCS mean field theory  \cite{BCS}. In general they have large superfluid density $n_s$,  (e.g. in two dimensions $\hbar^2 n_s/m >> T_c$), and weak phase fluctuations. 
 
  In contrast, under-doped  high $T_c$ cuprates\cite{Uemura,EK}, small capacitance Josephson arrays, and disordered thin films\cite{Kap},  are characterized by  low superfluid density. This enhances the role of phase fluctuations and  vortex delocalization near $T_c$.   Pairing correlations persist  well above $T_c$ \cite{Yazdani}, and  ''normal state''  transport coefficients\cite{Ong} do not follow familiar  Fermi liquid behavior.

To describe strongly fluctuating  superconductors,  it is natural to consider effective  Hamiltonians of charge $2e$ bosons\cite{PBFM,Qmelts}.
Continuum and weakly interacting superfluids are well approximated by the Gross-Pitaevski (GP) equation\cite{PS}.
 However, a strong periodic potential  generally enhance  the  short wavelength fluctuations beyond the validity of the GP approach.
Non uniform potentials are unavoidable in solid state superconductors. Nowadays,
their effects can be  systematically studied
in cold atom condensates on optical lattices\cite{SF-Mott}.
At strong interactions and commensurate fillings,  the superfluid is unstable toward  charge-gapped Mott insulator phases, or ''vortex condensates''\cite{FisherLee}.
The phase diagrams of lattice bosons have been studied extensively in recent years. However, little is known about their vortex dynamics and transport  coefficients, especially in the strongly interacting regime.

Here we report some recent results\cite{LAA} on  hard core bosons on finite toroidal clusters. We find several interesting  effects of the  lattice on the ground state  and Hall conductivity, which may have experimental implications.

\subsection{The Model}

The gauged quantum $XY$ model on a square lattice represents two dimensional hard core bosons in a perpendicular magnetic field:
\be
\cH = -t \sum_{\langle ij\rangle} \left(e^{i qA_{ij}} S^+_ i S^-_j + \mbox{H.c}  \right)   -  2\sum_i \mu_i S^z_i .
\label{QXY}
\ee
 The local density fluctuations are given by $n_i = S^z_i + \half$, and the superfluid order parameter is the magnetization in the $xy$ plane.
The mean field superfluid transition temperature goes as $T_c=t n_b(1-n_b)$, where $n_b=N_b/N$ is the filling fraction.
An important distinction between  hard core lattice bosons and
continuum bosons, is the existence of a charge conjugation symmetry
$C  \equiv  \exp\left(i\pi  \sum_\br S^x_\br \right)$.   $C$ transforms  ''particles'' into ''holes'', i.e.
 $n_i\to (1-n_i)$, and the Hamiltonian into
\be
C^\dagger \cH[\bA, n_b ]C  =   \cH[-\bA, 1-n_b ],
\label{Part-hole}
\ee
where $n_b=N_b/N$  is  the filling fraction.

%%%%%%%%%%%%%%%%%%%%%
\begin{figure}
\psfig{file=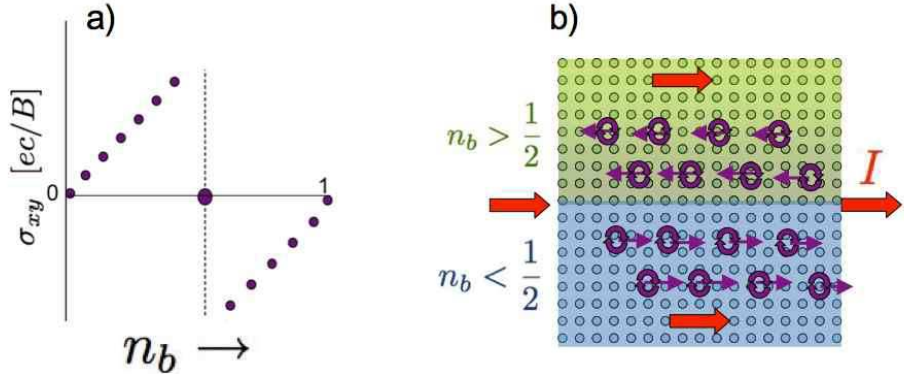,width=5in}
\caption{Reversal of Hall conductivity and Magnus action of hard core bosons at half filling. (a) The zero temperature Hall conductivity given
by the ground state Chern number of a 16 site square lattice on a
torus. (b)  Vortex drift directions (purple arrows)  in the
presence of a bias current (red arrows), for regions of lower
(blue) and higher (green) boson density than half filling.}
\label{fig:Hall-Magnus}
\end{figure}
%%%%%%%%%%%%%%%%%%%%
\subsection{Hall Conductivity} A consequence of
(\ref{Part-hole}) is that the Hall conductivity  is {\em
antisymmetric} in  $n_b-1/2$:
\be
\sigma_{H}(n_b,T)  = - \sigma_{H}(1-n_b,T).
\label{phs}
\ee
The temperature-dependent Hall
conductance of the finite cluster is given by the thermally
averaged  Chern numbers
\cite{yosi}.
A zero temperature  Hall conductance as a function of filling for
$N_\phi=1$ is plotted in Fig.~ \ref{fig:Hall-Magnus}a. At zero
temperature, $\sigma_H=N_b$ below half filling follows the
Galilean invariant result $\sxy
\propto N_b/N_{\rm v} $. At half
filling, $\sxy$ reverses sign  as expected by (\ref{phs}).

Fig.~\ref{fig:Hall-Magnus}a  shows a dramatic effect of the
lattice on the Hall coefficient:  $\sxy$ undergoes a sharp
transition between $\sxy>0$ ($\sxy <0$) just below (above) half
filling. For hard core bosons $\sxy(T,n_b)$ decreases with
temperature, with a characteristic temperature scale which
vanishes at half filling \cite{LAA}.

In terms of vortex dynamics, Hall conductivity inversion implies
that vortices suddenly drift in opposite directions as density of
bosons is varied near half filling (see
Fig.~\ref{fig:Hall-Magnus}b).   We propose to try to observe such
a dramatic effect for bosonic atoms on rotating optical lattices\cite{SF-Mott}.
As the density changes in space with the trapping potential,
vortices on either side of the half filling separatrix are
expected to flow in opposite directions relative the local
superflow.

%%%%%%%%%%%%%%%%%%%%%
\begin{figure}
\psfig{file=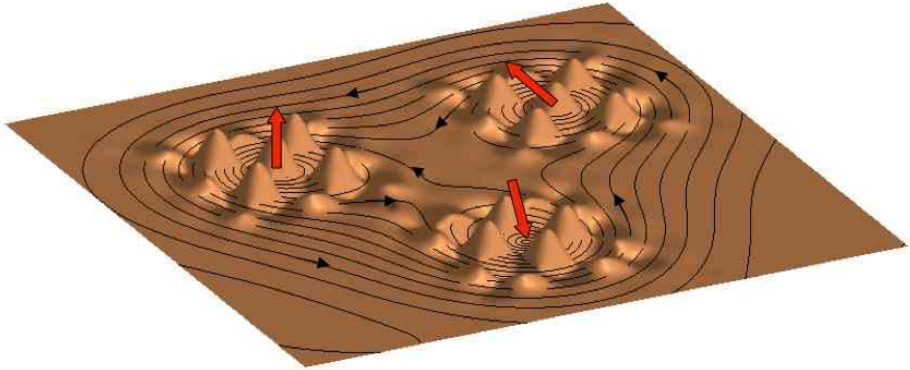,width=5in}
\caption{Illustration of three vortices of hard core bosons at half filling with charge density modulations in their cores.  Arrows depict directions of their v-spins.}
\label{fig:v-spin}
\end{figure}
%%%%%%%%%%%%%%%%%%%%

\subsection{V-spins of Vortices at Half Filling}
At half filling, vortices see no Magnus field, but instead they
acquire spin half quantum numbers we denote by {\em 'v-spins'}.
When the gauge field in  Eq. (\ref{QXY}) describes $N_\phi$ flux
quanta uniformly penetrating the torus, $N_\phi$ vortices are
inserted into the ground state. At half filling $n_b= 1/2$, for
any odd number of vortices $N_\phi=2 m+1, m=0,1,2,\ldots$, all
eigenstates are  at least two-fold degenerate. We have
proven\cite{LAA} that these doublets are associated with SU(2)
algebra of local symmetry operators. The v-spin in the ``$z$"
direction measures a  bipartite charge density wave
in the vortex core, as depicted in
Fig~\ref{fig:v-spin}. V-spin interactions between vortices decay
exponentially. V-spin excitations are expected to dominate the low
temperature thermodynamics at low values of external magnetic
field.

\subsection{Vortex Mass, and Vortex Lattice Melting}
At half filling, a vortex hops on the dual lattice with half a flux quantum per plaquette.
Its hopping rate  $t_{\rm v}$ was fit to exact numerical eigenenergies of  $\cH
$. Our results for  $N=20$,  show\cite{LAA} that  at half filling,  vortices
are as 'light' as bosons, $t_{\rm v}\approx t$.

When multiple vortices
are introduced by a magnetic field or rotation, they tend to localize in an Abrikosov
lattice which coexists with superfluidity. In two dimensions
the vortex lattice can melt by quantum  fluctuations resulting in a
non-superfluid Quantum Vortex Liquid (QVL).  A system of interacting vortices
can  be mapped to  the Boson Coloumb Liquid studied by Magro and
Ceperly (MC)
\cite{margo} . Using our
values of $t_{\rm v}$, the critical vortex melting  density was
bounded by  a surprisingly low vortex density,
 \be
n_{\rm v}^{\rm cr} \le \left(6.5-7.9 {V\over
t}\right)\times10^{-3} ~\mbox{vortices per site}.
\label{ncr}
 \ee
This implies that a
QVL is achievable at manageable rotation  frequencies for cold
atoms on optical lattices, and by moderate magnetic fields for Josephson junction arrays
and cuprate superconductors.

\end{document}